\documentstyle[12pt, a4]{article}

\begin{document}

\begin{flushright}
hep-th/9903135\\
NUP-A-98-8 \\
\end{flushright}
\hspace{\fill}

\vspace{14mm}
\begin{center}
\Large\bf{A Universal Lagrangian for Massive \\
Yang-Mills Theories without Higgs Bosons}
\end{center}
\hspace{\fill}
\begin{center}
{\large
Shinichi Deguchi \footnote{E-mail: deguchi@phys.cst.nihon-u.ac.jp}}
\end{center}

\begin{center}
Atomic Energy Research Institute,
College of Science and Technology, \\
Nihon University, Tokyo 101-8308, Japan
\end{center}

\hspace{\fill}
\vspace{5mm}
\begin{flushleft}
\bf{Abstract}
\end{flushleft}

A universal Lagrangian that defines various four-dimensional
massive Yang-Mills
theories without Higgs bosons is presented.
Each of the theories is characterized by a constant $k$ contained in the
Lagrangian.
For $k=0$, the Lagrangian reduces to one defining the topologically
massive Yang-Mills theory, and for $k=1$, the Lagrangian reduces to one
defining the Freedman-Townsend model.
New massive Yang-Mills theories are obtained by
choosing $k$ to be real numbers other than 0 and 1.

\vspace{4mm}


\newpage

Understanding of mass generation for gauge fields is one of the most
important subjects discussed in the non-Abelian gauge theories.
The Higgs model is widely accepted at present as a popular model describing
mass generation for Yang-Mills fields in four dimensions.
Although the Higgs model is attractive because of its renormalizable
structure, the Higgs bosons have not been found experimentally yet.
Hence, four-dimensional massive Yang-Mills theories without Higgs bosons
may be useful as alternative (or effective) theories of the Higgs model.

Until recently, several massive Yang-Mills theories in four dimensions
that involve no Higgs bosons have been presented.
The oldest theory is the non-Abelian St\"{u}ckelberg formalism [1]
or, equivalently, the gauged nonlinear sigma model [2].
The non-Abelian St\"{u}ckelberg formalism has actually been applied to
the construction of an electroweak model [3], and to the description of
gluon mass generation in continuum QCD [4].

Besides the non-Abelian St\"{u}ckelberg formalism,
a model presented by Freedman and Townsend [5] is known as a
four-dimensional massive Yang-Mills theory without Higgs bosons.
This model is constructed with a non-Abelian rank-2 antisymmetric
tensor field, which is treated in this model as an auxiliary field.
Freedman and Townsend showed the classical equivalence of their model to
the non-Abelian St\"{u}ckelberg formalism.

Recently, another four-dimensional massive Yang-Mills theory
without Higgs bosons
has been presented by Lahiri and by Barcelos-Neto, Cabo and Silva [6].
This theory is formulated as a non-Abelian generalization of
the topologically massive Abelian gauge theory (TMAGT) discussed in [7],
and will be called, in the present paper,
^^ ^^ topologically" massive Yang-Mills theory (TMYMT).
A non-Abelian rank-2 antisymmetric tensor field introduced in the TMYMT
has own kinetic term in the Lagrangian, and is not treated as an auxiliary
field.
A electroweak model based on this theory has also been considered by
Barcelos-Neto and Rabello [8].

In the present paper, we would like to propose a universal Lagrangian
that defines various four-dimensional massive Yang-Mills theories without
Higgs bosons.
We demonstrate that the Lagrangian for the TMYMT and that for
the Freedman-Townsend model
are obtained by choosing a constant contained in the universal
Lagrangian to be suitable numbers.
In addition to the TMYMT and the Freedman-Townsend model,
the universal Lagrangian defines
{\it new} massive Yang-Mills theories without Higgs bosons.

Let us begin by considering a four-dimensional non-Abelian gauge theory
defined by the Lagrangian
\begin{eqnarray}
{\cal L}_{(k)}&=&-{1\over4}{\rm Tr}(F_{\mu\nu}F^{\mu\nu}) \nonumber \\
& &+{m\over4}\epsilon^{\mu\nu\rho\sigma}
{\rm Tr} \{ B_{\mu\nu} ( F_{\rho\sigma}
-2D_{\rho}U_{\sigma}+2iqkU_{\rho}U_{\sigma})\}
\nonumber \\
& &-{i\over2}q\epsilon^{\mu\nu\rho\sigma}
{\rm Tr}(\phi_{\mu}[U_{\nu},\, F_{\rho\sigma}-2kD_{\rho}U_{\sigma}])
\nonumber \\
& &+{1\over2}m^{2}{\rm Tr}(U_{\mu}U^{\mu})
\end{eqnarray}
%
with
\begin{eqnarray}
D_{\mu} &\equiv& \partial_{\mu}+iq[A_{\mu},\;\;] \:, \\
F_{\mu\nu} &\equiv& \partial_{\mu}A_{\nu}-\partial_{\nu}A_{\mu}
+iq[A_{\mu},\,A_{\nu}] \:.
\end{eqnarray}
%
Here $A_{\mu}$ is a Yang-Mills field, $B_{\mu\nu}$ is a non-Abelian
antisymmetric tensor field, $\phi_{\mu}$ and $U_{\mu}$ are non-Abelian
vector fields, $m$ is a constant with dimensions of mass, and
$k$ and $q$ are dimensionless constants.
[$\,$The conventions for the metric signature and the Levi-Civita
symbol are $(+,-,-,-)$ and $\epsilon^{0123}=-1.\,$]
All the fields take values in the Lie algebra of a compact semisimple gauge
group $G$, and are expanded as $X=\sum_{a=1}^{\rm{dim}{\it G}}X^{a}T_{a}$
with respect to the generators $T_{a}$ of $G$.
Under the gauge transformation
\begin{eqnarray}
\delta A_{\mu}&=&D_{\mu}\lambda \:, \\
\delta B_{\mu\nu}&=&D_{\mu}\Lambda_{\nu}-D_{\nu}\Lambda_{\mu}
+iq[B_{\mu\nu},\, \lambda] \:, \\
\delta \phi_{\mu}&=&m\Lambda_{\mu}+iq[\phi_{\mu},\,\lambda] \:, \\
\delta U_{\mu}&=&iq[U_{\mu},\,\lambda] \:,
\end{eqnarray}
%
with the Lie algebra valued parameters $\lambda$ and
$\Lambda_{\mu}$,\footnotemark[1]
\footnotetext[1]{The vector field $\phi_{\mu}$ obeying the transformation
rule (6) has also been introduced in a Yang-Mills theory in loop space [9],
in which $\phi_{\mu}$ is treated as a constrained Nambu-Goldstone field on
loop space.}
the Lagrangian ${\cal L}_{(k)}$ remains invariant up to total derivative
terms. The Euler-Lagrange equations for $B_{\mu\nu}$, $\phi_{\mu}$ and
$U_{\mu}$ are derived from ${\cal L}_{(k)}$ as follows:
\begin{eqnarray}
\epsilon^{\mu\nu\rho\sigma}(F_{\rho\sigma}
-2D_{\rho}U_{\sigma}+2iqkU_{\rho}U_{\sigma})=0 \:, \\
\epsilon^{\mu\nu\rho\sigma}
[U_{\nu},\, F_{\rho\sigma}-2kD_{\rho}U_{\sigma}]=0 \:, \\
U^{\mu}={1\over6m}\epsilon^{\mu\nu\rho\sigma}
(\widehat{H}_{\nu\rho\sigma}+3iqk[\widehat{B}_{\nu\rho},\,U_{\sigma}]) \:,
\end{eqnarray}
%
where
\begin{eqnarray}
\widehat{B}_{\mu\nu}\equiv B_{\mu\nu}
-{1\over{m}}(D_{\mu}\phi_{\nu}-D_{\nu}\phi_{\mu}) \:,
\\
\widehat{H}_{\mu\nu\rho}\equiv D_{\mu}\widehat{B}_{\nu\rho}
+D_{\nu}\widehat{B}_{\rho\mu}+D_{\rho}\widehat{B}_{\mu\nu} \:.
\end{eqnarray}
%
We find from Eqs.(4), (5) and (6) that $\widehat{B}_{\mu\nu}$ and
$\widehat{H}_{\mu\nu\rho}$ transform gauge covariantly in the same way as
$U_{\mu}$.
Applying the covariant derivative $D_{\nu}$ on Eq.(8) and
summing over $\nu$ lead to Eq.(9); hence, Eq.(9) is compatible with Eq.(8).
The vector field $\phi_{\mu}$ is necessary to treat Eq.(9) as an
Euler-Lagrange
equation derived from ${\cal L}_{(k)}$.
Applying the covariant derivative $D_{\mu}$ on Eq.(9) and summing over $\mu$
do not yield further constraints by means of Eq.(8).

Before proceeding to discuss the non-Abelian theory, let us consider the
Abelian case $G=U(1)$.
The Lagrangian ${\cal L}_{(k)}$ then takes the following form:
\begin{eqnarray}
{\cal L}_{\rm A}&=&-{1\over4}F_{\mu\nu}F^{\mu\nu}
+{m\over4}\epsilon^{\mu\nu\rho\sigma}
B_{\mu\nu} ( F_{\rho\sigma}-2\partial_{\rho}U_{\sigma}) \nonumber \\
& &+{1\over2}m^{2}U_{\mu}U^{\mu} \:.
\end{eqnarray}
%
Here it should be noted that the Lagrangian ${\cal L}_{\rm A}$ does not
contain $\phi_{\mu}$ and $k$.
In this case, Eq.(8) becomes a simple equation
$\epsilon^{\mu\nu\rho\sigma}\partial_{\rho}(A_{\sigma}-U_{\sigma})=0$,
which can locally be solved as
\begin{eqnarray}
U_{\mu}=A_{\mu}-{1\over{m}}\partial_{\mu}\varphi
\end{eqnarray}
%
with a scalar field $\varphi$.
Substituting Eq.(14) into Eq.(13), we have the Lagrangian defining the
Abelian
St\"{u}ckelberg formalism:
\begin{equation}
{\cal L}_{\rm AS}=-{1\over4}F_{\mu\nu}F^{\mu\nu}
+{1\over2}(mA_{\mu}-\partial_{\mu}\varphi)
(mA^{\mu}-\partial^{\mu}\varphi) \:.
\end{equation}
%
On the other hand, Eq.(10) becomes
\begin{eqnarray}
U^{\mu}={1\over6m}\epsilon^{\mu\nu\rho\sigma}H_{\nu\rho\sigma} \:,
\end{eqnarray}
%
where $H_{\mu\nu\rho}\equiv \partial_{\mu}B_{\nu\rho}
+\partial_{\nu}B_{\rho\mu}
+\partial_{\rho}B_{\mu\nu}$. Substituting Eq.(16) into Eq.(13) and
removing a total derivative term, we obtain the Lagrangian defining the
TMAGT in four dimensions [7]:
\begin{equation}
{\cal L}_{\rm TA}=-{1\over4}F_{\mu\nu}F^{\mu\nu}
+{1\over12}H_{\mu\nu\rho}H^{\mu\nu\rho}
+{m\over4}\epsilon^{\mu\nu\rho\sigma}B_{\mu\nu}F_{\rho\sigma} \:.
\end{equation}
%
Thus, using an equation of motion, ${\cal L}_{\rm A}$ reduces
to ${\cal L}_{\rm AS}$ or ${\cal L}_{\rm TA}$;
the Lagrangians ${\cal L}_{\rm AS}$ and ${\cal L}_{\rm TA}$ are classically
equivalent. Their equivalence is also established at the quantum level by
using the path-integral method [10].
Equations (14) and (16) give the relation
$mA_{\mu}-\partial_{\mu}\varphi
={1\over6}\epsilon_{\mu\nu\rho\sigma}H^{\nu\rho\sigma}$,
which demonstrates the duality between $\varphi$ and $B_{\mu\nu}$.
The TMAGT is therefore dual to the Abelian St\"{u}ckelberg formalism both
at the classical and quantum levels.

We now return to the non-Abelian theory.
Apart from the trace over the generators $T_{a}$,
the Lagrangian ${\cal L}_{\rm A}$ is just the quadratic part of
${\cal L}_{(k)}$. From this we see that ${\cal L}_{(k)}$ describes
a massive Yang-Mills field with (bare) mass $m$.
Let us first consider the case $k=0$.
In this case, the right-hand side of Eq.(10) does not include $U_{\mu\,}$,
which makes possible to eliminate $U_{\mu}$ from the Lagrangian
${\cal L}_{(0)}$, given by Eq.(1) with $k=0$, so as to get a polynomial
Lagrangian containing no $U_{\mu\,}$.
Substituting $U^{\mu}={1\over6m}\epsilon^{\mu\nu\rho\sigma}
\widehat{H}_{\nu\rho\sigma}$
into ${\cal L}_{(0)}$ and removing a total derivative term, we obtain
\begin{eqnarray}
{\cal L}_{\rm T}&=&-{1\over4}{\rm Tr}(F_{\mu\nu}F^{\mu\nu})
+{1\over12}{\rm Tr}(\widehat{H}_{\mu\nu\rho}\widehat{H}^{\mu\nu\rho})
\nonumber \\
& &+{m\over4}\epsilon^{\mu\nu\rho\sigma}{\rm Tr}(B_{\mu\nu}F_{\rho\sigma})
\:.
\end{eqnarray}
%
The same is also obtained, quantum mechanically,
by carrying out the Gaussian integration over $U_{\mu}$
in the vacuum-to-vacuum amplitude based on ${\cal L}_{(0)}$.
Hence, ${\cal L}_{(0)}$ is equivalent to ${\cal L}_{\rm T}$
both at the classical and quantum levels.
The massive Yang-Mills theory defined by the Lagrangian ${\cal L}_{\rm T}$,
which we call {\it topologically massive Yang-Mills theory} (TMYMT),
has been presented by Lahiri and by Barcelos-Neto,
Cabo and Silva [6] as a non-Abelian version of the TMAGT discussed in [7].
The quantization of the TMYMT
has recently been studied by Lahiri and by Hwang and Lee [11].
A characteristic of ${\cal L}_{\rm T}$ is that it includes
a polynomial kinetic term for $B_{\mu\nu}$.
We would like to emphasize that unlike in the Abelian case, the vector field
$\phi_{\mu}$ is essential to the TMYMT to maintain the non-Abelian gauge
symmetry.

We next consider the case $k=1$. The Lagrangian ${\cal L}_{(1)}$,
given by Eq.(1) with $k=1$, can be written as
\begin{eqnarray}
{\cal L}_{\rm FT}&=&-{1\over4}{\rm Tr}(F_{\mu\nu}F^{\mu\nu})+{m\over4}
\epsilon^{\mu\nu\rho\sigma}{\rm Tr}(\widetilde{B}_{\mu\nu}f_{\rho\sigma})
\nonumber \\
& &+{1\over2}m^{2}{\rm Tr}(U_{\mu}U^{\mu}) \:,
\end{eqnarray}
%
where
\begin{equation}
\widetilde{B}_{\mu\nu}\equiv B_{\mu\nu}
-{iq\over m}([\phi_{\mu},\,U_{\nu}]-[\phi_{\nu},\,U_{\mu}])\:,
\end{equation}
%
and $f_{\mu\nu}$ is the field strength of
$V_{\mu}\equiv A_{\mu}-U_{\mu}\,$:
\begin{equation}
f_{\mu\nu}\equiv\partial_{\mu}V_{\nu}-\partial_{\nu}V_{\mu}
+iq[V_{\mu},\,V_{\nu}] \:.
\end{equation}
%
Notice that the second and third terms in ${\cal L}_{(1)}$ have simply
been expressed with the tensor field $\widetilde{B}_{\mu\nu}$;
the vector field $\phi_{\mu}$ hides in $\widetilde{B}_{\mu\nu}$, and
does not occur in ${\cal L}_{\rm FT}$ explicitly.
The massive Yang-Mills theory defined by the Lagrangian ${\cal L}_{\rm FT}$
has been presented by Freedman and Townsend [5], and
is called the {\it Freedman-Townsend model}.
The quantization of the  Freedman-Townsend model has been studied by
Thierry-Mieg and Baulieu in a systematic manner [12].
From Eqs.(4)--(7), the gauge transformation rules of
$\widetilde{B}_{\mu\nu}$ and $V_{\mu}$ are found to be
\begin{eqnarray}
\delta \widetilde{B}_{\mu\nu}&=&\nabla_{\mu}\Lambda_{\nu}
-\nabla_{\nu}\Lambda_{\mu}
+iq[\widetilde{B}_{\mu\nu},\, \lambda] \:,  \\
\delta V_{\mu}&=&\nabla_{\mu}\lambda \:,
\end{eqnarray}
%
where
\begin{eqnarray}
\nabla_{\mu}\equiv\partial_{\mu}+iq[V_{\mu},\;\;]\:.
\end{eqnarray}
%
Clearly, ${\cal L}_{\rm FT}$ is gauge-invariant up to a total derivative
term.
The Euler-Lagrange equation for $\widetilde{B}_{\mu\nu}$ is

\begin{equation}
f_{\mu\nu}=0 \:,
\end{equation}
%
which agrees with Eq.(8) with $k=1$.
In the case $k=1$, Eq.(9) reduces to the commutator of
Eq.(8) and $U_{\nu}$ with the summation over $\nu$, and hence is not
an independent equation of motion.
For this reason,
it is not necessary to take into account Eq.(9) with $k=1$;
the vector field $\phi_{\mu}$ is not essential
to the Freedman-Townsend model.
If $U_{\mu}$ is eliminated from ${\cal L}_{\rm FT}$ by repeatedly using
Eq.(10) with $k=1$,
then ${\cal L}_{\rm FT}$ turns out to be a Lagrangian including
a nonpolynomial kinetic term for $B_{\mu\nu}$.
This shows a difference between the TMYMT and the Freedman-Townsend model.

The equation (25) can locally be solved as
\begin{eqnarray}
V_{\mu}={1\over{iq}}v^{-1}\partial_{\mu}v
\end{eqnarray}
%
in terms of $v(\in G)$ represented as
$v=\exp\!\left({iq\over{m}}\sum_{a}\varphi^{a}T_{a}\right)$
with scalar fields $\varphi^{a}$.
After substitution of Eq.(26) into Eq.(19),
we have the Lagrangian defining the non-Abelian St\"{u}ckelberg
formalism [1,2,17]:
\begin{eqnarray}
{\cal L}_{\rm S}&=&-{1\over4}{\rm Tr}(F_{\mu\nu}F^{\mu\nu})
+{1\over2}m^{2}{\rm Tr}(\widetilde{U}_{\mu}\widetilde{U}^{\mu}) \:,
\end{eqnarray}
%
where $\widetilde{U}_{\mu}\equiv A_{\mu}-{1\over{iq}}v^{-1}\partial_{\mu}v$.
The Freedman-Townsend model is therefore classically equivalent to
the non-Abelian St\"{u}ckelberg formalism [5].
Since the TMYMT is classically different from the Freedman-Townsend model,
the TMYMT can not be considered, at least classically, the dual theory
of the non-Abelian St\"{u}ckelberg formalism.
[$\,$In the Abelian limit $q\rightarrow0$,
the TMYMT becomes dual to the St\"{u}ckelberg formalism
both at the classical and quantum levels.$\,$]
The classically dual theory of the non-Abelian St\"{u}ckelberg formalism is
described by the nonpolynomial Lagrangian
that is obtained by
eliminating $U_{\mu}$ from ${\cal L}_{\rm FT}$ by repeatedly using
Eq.(10) with $k=1$ expressed as
\begin{eqnarray}
U^{\mu}={1\over{2m}}\epsilon^{\mu\nu\rho\sigma}\nabla_{\nu}
\widetilde{B}_{\rho\sigma} \:.
\end{eqnarray}
%
The nonpolynomial Lagrangian is thus written in terms of
$A_{\mu}$ and $\widetilde{B}_{\mu\nu}$.
The duality between $\varphi(\equiv\varphi^{a}T_{a})$ and
$\widetilde{B}_{\mu\nu}$ is guaranteed from the relation
$\widetilde{U}^{\mu}={1\over{2m}}\epsilon^{\mu\nu\rho\sigma}
(\partial_{\nu}\widetilde{B}_{\rho\sigma}+\cdots)$ whose right-hand side
is a power series that is determined from Eq.(28) by iteration with respect
to $U_{\mu}$.

The equivalence and the difference between the massive Yang-Mills theories
characterized by ${\cal L}_{\rm T}$, ${\cal L}_{\rm FT}$, and
${\cal L}_{\rm S}$ have been discussed at the classical level with aid of
the equations of motion.
In general, classical equivalence between two theories does not imply
their equivalence at the quantum level [13], and vice versa [14].
Since the Lagrangians ${\cal L}_{\rm T}$, ${\cal L}_{\rm FT}$,
and ${\cal L}_{\rm S}$ involve interaction terms, that is,
the terms disappearing in the Abelian limit,
it is by no means clear whether
the equivalence and the difference discussed above persist
at the quantum level.
Further study is needed to understand quantum-mechanical
equivalence and difference between the massive Yang-Mills theories.

The Lagrangian ${\cal L}_{(k)}$ can collectively define various
four-dimensional massive Yang-Mills theories distinguished by
the constant $k$. Among them, the TMYMT and the Freedman-Townsend model,
which correspond to the cases $k=0$ and $k=1$ respectively,
are considered particular theories.
We would like to emphasize in this paper that
besides these known theories,
there exist {\it new} massive Yang-Mills theories obtained by choosing $k$
to be real numbers other than 0 and 1.
It should be noted that the new theories need both the vector fields
$\phi_{\mu}$ and $U_{\mu}$: $\phi_{\mu}$ and $U_{\mu}$ are necessary
for ${\cal L}_{(k)}$ to be gauge invariant up to total derivative terms
and to be polynomial, respectively.
Since ${\cal L}_{(k)}$ describes various four-dimensional massive
Yang-Mills theories without Higgs bosons,
we may call ${\cal L}_{(k)}$ a ^^ ^^ universal" Lagrangian for
massive Yang-Mills theories without Higgs bosons.
In the Abelian limit, ${\cal L}_{(k)}$ becomes independent of $k$;
that is, all the massive Yang-Mills theories defined by ${\cal L}_{(k)}$
have
the same Abelian limit.
Apart from the trace over the generators $T_{a}$, this limit can be regarded
as the TMAGT because of the equivalence of ${\cal L}_{\rm A}$ and
${\cal L}_{\rm TA}$.

Non-Abelian generalizations of the TMAGT have also been
studied in terms of the antifield-BRST method without introducing extra
fields such as $\phi_{\mu}$ and $U_{\mu}$ [15].
This study concludes that it is not possible to
generalize ${\cal L}_{\rm TA}$ so as to incorporate power-counting
renormalizable interaction terms consisting only of $A_{\mu}$ and
$B_{\mu\nu}$. Obviously the non-Abelian generalizations defined by
${\cal L}_{(k)}$ lie outside the scope of the proof in Ref.[15].

Needless to say,
it is important to investigate the renormalizability and the unitarity of
the massive Yang-Mills theories defined by ${\cal L}_{(k)}$.
Beginning the investigation, we, however, face a difficulty at once:
the vector field $\phi_{\mu}$ has no kinetic term in ${\cal L}_{(k)}$, and,
in the Abelian limit, disappears from ${\cal L}_{(k)}$.
Consequently, the ordinary perturbative procedures can not be used to
investigate the renormalizability of the theories.
In addition, the proof of unitarity based on the Kugo-Ojima quartet
mechanism [16] can not be applied to the theories.
The point we have to consider first is therefore
how we should treat $\phi_{\mu}$ within the framework of the perturbation
theory.

As for the Freedman-Townsend model, the problem of disappearing $\phi_{\mu}$
can be avoided by treating $\widetilde{B}_{\mu\nu}$ as a fundamental
field. The unitarity of the Freedman-Townsend model is then shown in terms
of
the Kugo-Ojima quartet mechanism, while
the renormalizability of the model is still an open question;
it is by no means obvious whether the perturbatively nonrenormalizable
structure of the non-Abelian St\"{u}ckelberg formalism
in four dimensions [17] persists in the Freedman-Townsend model,
because their equivalence is not clear at the quantum level.
Thus, the renormalizability of the Freedman-Townsend model should also be
studied in the future.

\vspace{4mm}

\noindent
{\bf Acknowledgements}

\vspace{2mm}

I am grateful to Professor S. Naka and other members of the Theoretical
Physics Group at Nihon University for their encouragements and useful
comments. This work was supported in part by the Nihon University
Research Grant.

\newpage

\begin{center}
{\Large\bf References}

\vspace{1mm}

\end{center}
\begin{enumerate}


\item
T. Kunimasa and T. Got\={o}, Prog. Theor. Phys. {\bf 37}, 452 (1967).
%
\item
A. Salam and J. Strathdee, Phys. Rev. {\bf 184}, 1750 (1969).
%
\item
T. Sonoda and S. Y. Tsai, Prog. Theor. Phys. {\bf 71}, 878 (1984).
%
\item
J. M. Cornwall, Phys. Rev. {\bf 26}, 1453 (1982).
%
\item
D. Z. Freedman and P. K. Townsend, Nucl. Phys. {\bf B177}, 282 (1981).
%
\item
A. Lahiri, ^^ ^^ Generating vector boson masses", hep-th/9301060;
J. Barcelos-Neto, A. Cabo and M. B. D. Silva, Z. Phys. {\bf C72}, 345
(1996).
%
\item
E. Cremmer and J. Scherk, Nucl. Phys. {\bf B72}, 117 (1974);
A. Aurilia and Y. Takahashi, Prog. Theor. Phys. {\bf 66}, 693 (1981);
I. Oda and S. Yahikozawa, {\it ibid}.  {\bf 83}, 991 (1990);
T. J. Allen, M. J. Bowick and A. Lahiri, Mod. Phys. Lett. {\bf A6},
559 (1991).

%
\item
J. Barcelos-Neto and  S. Rabello, Z. Phys. {\bf C74}, 715 (1997).
%
\item
S. Deguchi and T. Nakajima, Int. J. Mod. Phys. {\bf A10}, 1019 (1995).
%
\item
S. Deguchi, T. Mukai and T. Nakajima, Phys. Rev. {\bf D59}, 065003 (1999).
%
\item
A. Lahiri, Phys. Rev. {\bf D55}, 5045 (1997);
D. S. Hwang and C. Y. Lee, J. Math. Phys. {\bf 38}, 30 (1997).
%
\item
J. Thierry-Mieg and L. Baulieu, Nucl. Phys. {\bf B228}, 259 (1983);
J. Thierry-Mieg, {\it ibid}. {\bf B335}, 334 (1990).
%
\item
C. R. Nappi, Phys. Rev. {\bf D21}, 418 (1980).
%
\item
S. Coleman, Phys. Rev. {\bf D11}, 2088 (1975); S. Mandelstam, {\it ibid}.
{\bf D11}, 3026 (1975).
%
\item
M. Henneaux, V. E. R. Lemes, C. A. G. Sasaki, S. P. Sorella, O. S. Ventura
and L. C. Q. Vilar, Phys. Lett. {\bf B410}, 195 (1997).
%
\item
T. Kugo and I. Ojima, Prog. Theor. Phys. Suppl. No66, 1 (1979);
N. Nakanishi and I. Ojima, {\it Covariant Operator Formalism of
Gauge Theories and Quantum Gravity} (World Scientific, Singapore, 1990)
%
\item
K. Shizuya, Nucl. Phys. {\bf B87}, 255 (1975); {\bf B94}, 260 (1975);
{\bf B121}, 125 (1977);
W. A. Bardeen and K. Shizuya, Phys. Rev. {\bf D18}, 1969 (1978);
Yu. N. Kafiev, Nucl. Phys. {\bf B201}, 341 (1982).

\end{enumerate}

\end{document}